\documentclass{elsart4}

\usepackage{amssymb}

\usepackage{graphics}
\usepackage{graphicx}
\usepackage{epsfig}
\usepackage{amssymb}
\journal{Physica B}

\usepackage[T1]{fontenc}
\usepackage[latin9]{inputenc}
\usepackage{color}
\usepackage{float}
\usepackage{amsmath}
\usepackage{graphicx}
\usepackage{amssymb}
\usepackage{esint}

\begin{document}

\begin{frontmatter}

% Title, authors and addresses

% use the thanksref command within \title, \author or \address for footnotes;
% use the corauthref command within \author for corresponding author footnotes;
% use the ead command for the email address,
% and the form \ead[url] for the home page:
 \title{Geometrically-Frustrated Pseudogap Phase of Coulomb Liquids}

 \author{Y. Pramudya,H. Terletska, S. Pankov, E. Manousakis, V. Dobrosavljevi\'c}

 \address{National High Magnetic Field Laboratory-Florida State University, 1800 E. Paul Dirac drive, Tallahassee, FL 32310,USA}
 
% \ead{email address}
% \ead[url]{home page}

% \address{Address\thanksref{label3}}
% \thanks[label3]{}

%\title{}

% use optional labels to link authors explicitly to addresses:
% \author[label1,label2]{}
% \address[label1]{}
% \address[label2]{}

%\author{}

%\address{}

\begin{abstract}
% Text of abstract
We study a class of models with long-range repulsive interactions of the generalized Coulomb form $V(r)\sim 1/r^{\alpha}$. We show that decreasing the interaction exponent in the regime $\alpha < d$  dramatically depresses the charge ordering temperature $T_c$ in any dimension $d\ge 2$, reflecting the strong geometric frustration produced by long-range interactions. A nearly frozen Coulomb liquid then survives in a broad pseudogap phase found at $T > T_c$, which is characterized by an unusual temperature dependence of all quantities. In contrast, the leading critical behavior very close to the charge-ordering temperature remains identical as in models with short-range interactions. 
\end{abstract}

\begin{keyword}
% keywords here, in the form: keyword \sep keyword
pseudogap, Coulomb, Universality Class, EDMFT 
% PACS codes here, in the form: \PACS code \sep code
\PACS{71.30.+h,71.27.+a}
\end{keyword}
\end{frontmatter}

% main text
\section{Introduction}
\label{}

The phenomenon of screening has long been known in presence of long-range Coulomb interactions, and it is generally expected to render the observed behavior very similar to that of systems with short-range interactions. Indeed, recent computational work has investigated the critical behavior close to charge ordering in lattice Coulomb systems \cite{Mobius2009}, suggesting the Ising universality class. In addition, analytical arguments have been presented \cite{Troster2010} supporting this view, for a broad class of lattice models with the generalized Coulomb interaction of the form $V(r) \sim 1/r^{\alpha}$ in $d\ge 2$ dimensions. We should mention, however, that long-range interactions are generally expected to produce mean-field like critical behavior for $d < \alpha < d^*$, while for  $d < d^* < \alpha $ one expect short-range critical behavior. The nontrivial effect of long-range interactions upon the critical behavior is possible only for $\alpha > d$, because in this regime the interaction is ``integrable'' (no neutralizing background is needed), and screening becomes inoperative. 

Should one expect any interesting or novel physics in Coulomb-like models ($\alpha < d$), in comparison to  the short-range situation? Our work confirms  that these models indeed feature conventional critical behavior in the narrow critical region $T\approx T_c$. We show, however, that a striking new behavior is uncovered in a broader temperature interval, reflecting strong geometric frustration inherent to such long-range interactions. First, we find a dramatic decrease of melting temperature of lattice Coulomb gas as a result of the level of frustration in the system. This can be easily understood by noting that our lattice Coulomb gas maps into an antiferromagnetic Ising model with long-range interactions. Here, all the spin tend to anti-align with all other spins, but this cannot be achieved for very long interactions ($\alpha < d$), resulting in very low melting temperature. Indeed, for half-filled Coulomb systems ($\alpha = 1$), the melting temperature $T_{c}$ is one order of magnitude smaller \cite{efros92} than the generalized Coulomb energy $E_{c}=e^{2}/a$ , where $a$ is the lattice constant. Continuum models \cite{Wigner34,tanatar89prl,ramirez94arms} show an even more dramatic behavior, with the melting temperature being as much as two orders of magnitude smaller then the Coulomb energy. This striking behavior, although well documented in several model studies, is not widely appreciated or understood in simple terms.

The second robust feature of these models, which has only recently been discovered  \cite{yohanesprb2011}, is the emergence of the ``pseudogap phase''. This pseudogap phase is a specific feature of long-range type interaction with $\alpha < d$, and is observed in a broad temperature range $T_{c }< T < T^*$  (see Fig. 1), where $T^*$ is the pseudogap temperature where the gap in the single particle density of state (DOS) starts to open.  We show that the physical Coulomb interaction ($\alpha = 1$) lays deep in the regime of very long range interaction $\alpha\rightarrow 0$, the regime where  our analytical ``Extended Dynamical Mean-Field Theory'' (EDMFT) approach becomes is asymptotically exact  \cite{yohanesprb2011}. This observation explains the surprising accuracy of this analytical scheme both when applied to clean models  \cite{yohanesprb2011}, and in previous applications to Coulomb glasses \cite{pankov05prl}.

\begin{figure}[b]
 \vspace{0.2cm}
\includegraphics[width=7.5cm]{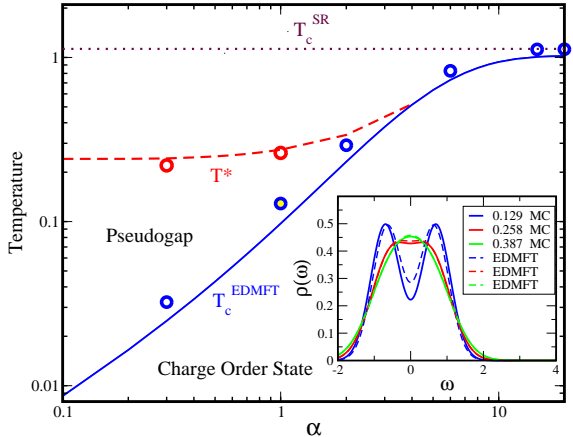}

\caption{(Color online) Phase diagram of the half-filled classical $D=3$ lattice
model with interactions $V(R)=R^{-\alpha}$. The charge ordering temperature
$T_{c}(\alpha)\sim\alpha$, as obtained from EDMFT theory (full line)
and Monte-Carlo simulations (open symbols). The pseudogap temperature
$T^{*}$(dashed line) remains finite as $\alpha\rightarrow0$; a broad
pseudogap phase emerges at $\alpha\le d$. We also show $T_{c}^{SR}\approx1$
for the same model with short-range interactions (dotted line). 
The inset shows corresponding single particle density of state (DOS) for different
temperatures. The EDMFT results (dashed line) show excellent agreement with Monte-Carlo 
simulation (solid line) above the melting temperature $T_c$.  \cite{yohanesprb2011,pankov05prl} }

\end{figure}

\section{Model and EDMFT Approach}

Our model is described by the Hamiltonian of particles living on a half-filled hypercube lattice 
(lattice spacing $a$) that interact via Coulomb-like interaction $V(R_{ij})\sim(R_{ij}/a)^{-\alpha}$,
\begin{equation}
H=\frac{1}{2}\sum_{ij}V(R_{ij})(n_{i}-\langle n\rangle)(n_{j}-\langle n\rangle).\label{hamiltonian}\end{equation}
Here $R_{ij}$ is the distance between
lattice sites $i$ and $j$ expressed in the units of the lattice
spacing. In general, our unit of energy the nearest-neighbor repulsion $E_c = V(1)$ . We focus on the half-filled system $<n>=1/2$
that maps the problem into {}``antiferromagnetic-like'' Ising model. In addition to numerically-exact (classical) Monte-Carlo simulations, we also utilize an analytical approach, called so-called Extended Dynamical Mean Field Theory (EDMFT)  \cite{pankov-motome,chitra00prl} which is expected  \cite{yohanesprb2011} to be accurate for very long interaction $(\alpha\ll1)$, where the effective coordination number becomes very large. The
inter-site charge correlations are included in this approach, and are treated on the same footing as the local ones \cite{chitra00prl}.  Here, the lattice problem is mapped into an effective impurity problem embedded in a self-consistently determined bath containing both fermionic and bosonic excitations \cite{pankov-motome,chitra00prl}.  In our case, this bosonic bath describes the plasmon excitations induced by the inter-site charge correlations. 

\begin{figure}[t]
\includegraphics[width=2.9in]{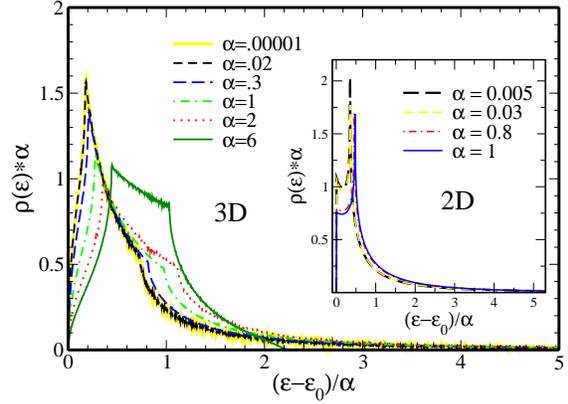}\caption{(Color online) Scaling behavior of the spectral density $\nu(\varepsilon)$ of plasmon modes on a 3D hypercubic lattice.  The low-energy branch of the spectrum, which describes the "sheer" charge fluctuations close to the ordering wavevector, features an $\alpha$-dependent energy scale, one that sets the melting temperature $T_c \sim \alpha$.  The scaling becomes exact for very long interaction range $\alpha \ll 1$, but is approximately valid for all $\alpha \le 1$, explaining the simple $\alpha$-dependence of all quantities.}

\end{figure}

In this paper, we focus on the classical limit, where the
origin of the pseudogap phase is most pronounced and the accuracy
of the EDMFT can be tested directly with simple classical Monte
Carlo simulations. Here, the density-density correlation function becomes 
\begin{equation}\chi(k)=(4+D+\beta V_{k})^{-1},\end{equation}
and the self-consistent condition \begin{equation}
\frac{1}{4}=\int d\varepsilon\nu(\varepsilon)(4+D+\beta\varepsilon)^{-1},\end{equation}
where we define the classical plasmon-mode spectral density (Fig. 2)
 %\begin{equation}
 $\nu(\varepsilon)=\sum_{k}\delta(\varepsilon-V_{k}).$
 %\end{equation}  
By solving numerically the self-consistent condition in equation (3) using the spectral density $\nu(\varepsilon)$,
the $T$ dependence of dimensionless parameter $D(T)$ is found to be
power-law dependent (Fig. 3).  The analytic solution can be derived at the limit of $\alpha\ll d$, spectral density becomes 
$\nu(\varepsilon)\backsimeq \delta(\varepsilon-\varepsilon_{0})$, where  
$\varepsilon_{0}=V_{Q}$ ($Q=(\pi,\pi)$ in 2D, and  $Q=(\pi,\pi,\pi)$ in 3D)  
is the minimum energy in momentum space (center of Brillouin zone). 
By solving equation (2) at this limit, we find the analytic solution 
%\begin{equation}
$D(T)=T^{-1}V_{Q}.$
%\end{equation} 
The result qualitatively holds all the way to $\alpha \sim d$ as shown in Fig. 3.  

The charge ordering temperature $T_{c}(\alpha)$ is defined
by the $\chi(k)\rightarrow\infty$ at the corresponding ordering wave
vector $k=Q$; in $d=3$ the system forms a BCC structure, and a checkerboard patter in $d=2$.
The depression of melting temperature  as a function of interaction parameter $\alpha\ll1$ can be understood
in simple way by noting that the spectral density $\nu(\varepsilon)$
has a simple scaling form \begin{equation}\nu(\varepsilon)=\alpha^{-1}\widetilde{\nu}\{(\varepsilon-\varepsilon_{0})/\alpha\},\end{equation}
as shown in Fig. 2. The sharp peak of $(\varepsilon-\varepsilon_{0})$
below characteristic energy scale $\varepsilon^{*}(\alpha)\sim\alpha$,
physically correspond to {}``sheer'' plasmon modes with wave vector
$k\approx Q$. \cite{tanatar89prl,BonsalMaradudin} This energy scale plays
a role of an effective Debye temperature. This tells us why the ordering
temperature decreases  \begin{equation}T_{c}(\alpha)=\alpha\int d\varepsilon\tilde{\nu}(\varepsilon)/\varepsilon\sim\varepsilon^{*}(\alpha)\sim0.1\alpha,\end{equation}
in agreement with the estimation of Lindermann criterion applied to
the sheer mode. At the lowest energy near $\varepsilon_{0}$ the dispersion
relation assumes the standard form $\nu(\varepsilon)\sim\varepsilon^{(d-2)/2}$, same as for short-range interactions.  This can also be understood
from the form of the potential surface in momentum space, as shown in Fig. 4.  The potential surface is  flatter for smaller
$\alpha$ and the values are close to the minimum energy level at $Q=(\pi, \pi)$.  This picture clearly explains
the origin of the sharp peak in the spectral density $\nu(\varepsilon)$.

\begin{figure}[t]
 \vspace{0.1cm}

\includegraphics[width=2.9in]{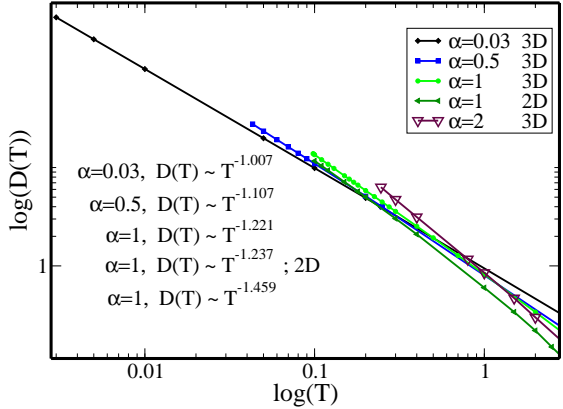}\caption{(Color online) The power law behavior of $D(T)$
 for different interaction range $\alpha $.  
In the limit of $\alpha \rightarrow 0$, we find parameter $D(T)\sim T^{-1}$, both in d=2 and d=3.}

\end{figure}

In the classical limit, the single particle density of state (DOS) is
nothing more than the probability distribution function of electrostatic potentials
computed on different sites in the lattice 

\begin{equation}\rho(\omega,T)\equiv\left\langle \sum_{i}\delta(\omega-\sum_{j}V_{ij}n_{j})\right\rangle.
\end{equation}
Within the EDMFT approach it assumes the form
\begin{multline}
\rho(\omega,T)=\frac{\beta}{\sqrt{8\pi D}}\left\{ \exp{\left[-\frac{\beta^{2}}{2D}\left(\omega+\frac{D}{2\beta}\right)^{2}\right]}\right.\\
\left.+\exp{\left[-\frac{\beta^{2}}{2D}\left(\omega-\frac{D}{2\beta}\right)^{2}\right]}\right\} .\label{classdos}\end{multline}

The Coulomb pseudogap ({}``plasma dip'') $E_{gap}=D/\beta$
, which is the distance between the Gaussian peaks that start to open
at $D=E_{gap}/4$ . As shown in Fig.1, both $E_{gap}$ and $T^{*}$
remain finite for $\alpha\ll1$, since $D(T)\thickapprox\beta$ in
this limit (Fig. 3). This results, for long range interactions, in the formation of a wide pseudogap
phase, since the melting temperature decreases as $T_{c}(\alpha)\sim0.1\alpha$
while pseudogap temperature remains finite ($T^* \rightarrow 1/4$) as $\alpha \rightarrow 0$. 
We define $T^*$ based on the change of curvature in the DOS at $\omega=0$, i.e. the vanishing of the second derivative of the DOS
$d^2\rho(\omega)/d\omega=0\rightarrow D=8$, and finally use $D(T^*)=8$ from the power-law dependent of $D(T)$ (Fig. 3).

\begin{figure}[b]
 \vspace{0.1cm}

\includegraphics[width=2.9in]{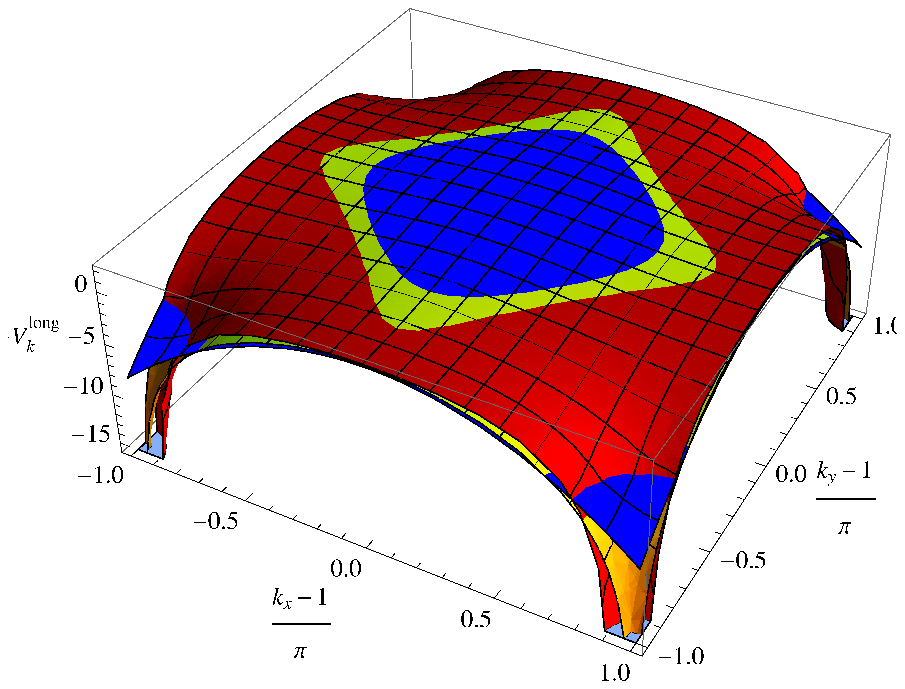}
\includegraphics[width=2.9in]{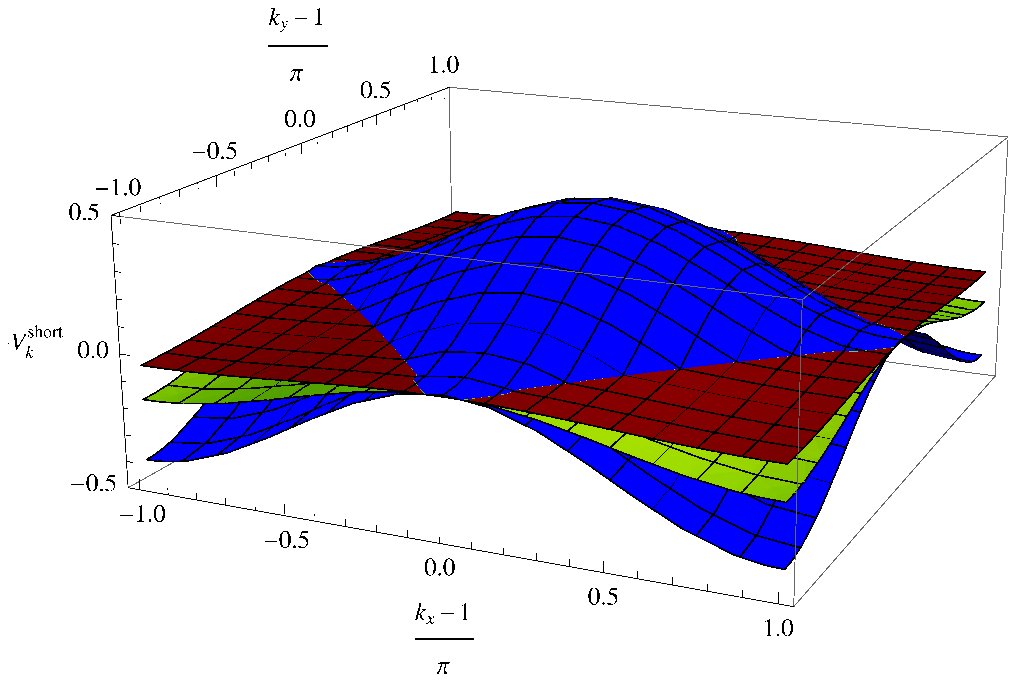}

\caption{(Color online) (a) The dispersion of the long-range part of Coulomb potential in $d=2$ ($J_0=-1$ [reverse potential surface]) according to equation (8,9) for different interaction range.  The lowest energy state  near $Q=(\pi,\pi)$  feature parabolic like surface for  interactions $\alpha$ equal to 1 (red/flattest),2(green/middle),3(blue/curviest) and only change rapidly near the edges of the first Brillouin zone.   The edges are related to the tail on the plasmon spectral density of Fig. 2 and the parabolic surface related to the peak on the plasmon spectral density.  At $\alpha>d$, the long-range dispersion expansion does not have rapid change near the edges and is more like short-range dispersion. (b) Short-range dispersion expansion does not have singularity near the edges for any interaction range. }  

\end{figure}

\section{Long Range Interaction: Universality Class}

The fact that models with long-range interactions with $\alpha < d$ belong to the same universality class as the short-range 
Ising model can be clearly seen from carefully examining the form of the interaction in momentum space. To determine its precise form in quantitative detail one must use Ewald summation \cite{ewald,ewald-jchem,ewald-smith}
and split the appropriate  lattice into a short-range part 
\begin{equation}V_{k}^{short}=J_0\sum_{r_{ij}\ne0}e^{i(Q+k)r_{ij}}\frac{\Gamma(\frac{\alpha}{2},|r_{ij}|^{2})}{\Gamma(\frac{\alpha}{2})},\end{equation}
and long-range part
\begin{multline}
V_{k}^{long}=J_0\frac{\pi^{\alpha-d/2}}{L^{d}}\sum_{G\in\mathbb{Z}^{*}}\left\{ \left\{ \frac{(Q+k+G)}{2\pi}\right\} ^{\alpha-d}\right.\\
\left.\frac{\Gamma(\frac{d-\alpha}{2},|\frac{(Q+k+G)}{2\pi}|^{2})}{\Gamma(\frac{\alpha}{2})}\right\} -\frac{2\pi^{\alpha/2}}{\alpha\Gamma(\alpha/2)},\label{eq:9}\end{multline}
where $\Gamma(a,x)$  is the incomplete Gamma function, and $G$ is reciprocal lattice 
unit vector.  

The resulting dispersion is shown, for different $\alpha$ in Fig. 4. At the center of the Brillouin zone, the long-range
part for different interaction $\alpha$ is equally parabolic for
all k vectors and only start to drop sharply when approaching the
vicinity of the corners of momentum space $\mathbb{Z}^{*}$. The singularity points near the edges
of the Brillouin zone where $k_i=2\pi \mathbb{Z}$ diverges to  $V_k\rightarrow \infty $, leading to
positively diverging energy for corresponding (longitudinal) fluctuation modes. This mode has very small statistical weight. It is related to the tail on the spectral density (Fig. 2), and it does not contribute in the critical behavior of the system. The critical behavior, in contrast, is determine by the dispersion of ``sheer''  modes in the vicinity of the ordering wavevector, a quantity that remains of the same qualitative form as for short-range interactions \cite{Troster2010}.

\section{Conclusions}

The role of long-range interaction in suppressing critical temperature $T_c$ is an important feature of systems with long-range interactions, behavior which directly reflects the generic presence of strong geometric frustration. While these effects do not modify the universality class of the narrow critical region, novel behavior is found well above the melting temperature. Here we uncovered the emergence of a broad pseudogap phase, a feature unique to long-range interactions; this regime is very accurately described by our analytical EDMFT theory, and we find excellent agreement with (numerically exact) Monte Carlo simulations. 

\section{Acknowledgements}

The authors thank Seng Cheong, Misha Fogler, Daniel Khomskii, Andy
Millis, Joerg Schmalian, Dan Tsui, and Kun Yang for useful discussions, and also
to the organizers especially Natasha Kirova who incorporated all of our interest. 
This work was supported by the National High Magnetic Field Laboratory
(YP, HT, SP, EM, and VD) and the NSF through grants DMR-0542026 and
DMR-1005751 (YP, HT, and VD).

% The Appendices part is started with the command \appendix;
% appendix sections are then done as normal sections
% \appendix

% \section{}
% \label{}

\end{document}